# Meeting Student Needs for Multivariate Data Analysis: A Case Study in Teaching a Multivariate Data Analysis Course with No Pre-requisites

Amy Wagaman, Amherst College

## Abstract

Modern students encounter big, messy data sets long before setting foot in our classrooms. Many of our students need to develop skills in exploratory data analysis and multivariate analysis techniques for their jobs after college, but these topics are not covered in introductory statistics courses. This case study describes my experience in designing and teaching a course on multivariate data analysis with no pre-requisites, using real data, active learning, and other activities to help students tackle the material.

## 1. Introduction

Modern students are encountering and using data in way unlike their past counterparts. Their first experience with data is not in statistics courses and the type of data they encounter (big, messy, complex data sets) is not usually part of introductory statistics courses (Gould 2010). When our students graduate and enter the workforce, many of their jobs include data management (retrieving, filtering, or cleaning) and performing basic exploratory data analysis (Nolan and Temple Lang 2010). Yet, these topics are not major foci of our introductory statistics courses.

Introductory statistics courses are undergoing significant change. We are moving away from the tyranny of the computable (Cobb 2007) and have shifted towards using randomization based procedures to introduce inference (Tintle et al. 2011). However, we still are not meeting the needs of many of our students in terms of dealing with big data sets, visualizing data, and doing exploratory data analysis. This leads to some serious questions.
- (Q1) How can we introduce our students to big data sets and basic techniques for multivariate data analysis when those students have little or no background in statistics? Can and should we include this in introductory statistics courses?
- What do we want students to get out of a course designed to address question 1?
- What teaching approaches/principles will work to help students learn these techniques?
- What computational skills can we help our students develop in such a course?

Adding multivariate analysis topics to introductory statistics would strain an already taxed curriculum. Instead, the first question can be addressed by developing a course to introduce students to big data and appropriate analysis techniques, and not include these topics in introductory statistics courses. This new course should have minimal or no pre-requisites in order to have a broad student audience. With this in mind, it is quickly clear that we are not talking about developing a second course in statistics (i.e. something to follow introductory statistics). Second courses can be regression-intensive, cover analysis of variance, touch on design, or introduce categorical data analysis, all relying on knowledge students gained in introductory statistics. There is some development of a second course along these lines, especially as a first course after Advanced Placement (AP) statistics (Kuiper and Sklar 2012). A course designed to address the first question though, needs to have exploratory data analysis and techniques like principal components analysis (PCA) and clustering methods, without relying on a semester's worth of introductory statistics knowledge. You may wonder if this is even possible. I believe the answer is yes, although it depends on what material you hold the students accountable for. This case study describes a course I designed to address these questions.

Thinking through the questions, once I decided to design a course to address the first, I needed to set course goals. I decided that the overall course goals were for students to be introduced to big data sets, and gain an applied understanding of exploratory data analysis and several multivariate analysis techniques. That meant, for each technique, they should be able to recognize appropriate and inappropriate applications, be able to interpret results, including justifying choices they had to make during the analysis, and communicate their results. In order to accomplish this, it was clear that students would need to see many examples, have time in the computer lab to practice on their own, and receive feedback on their communication of the ideas. By making the course very applied, instead of theoretical, I would not need to rely on student background knowledge, and taught the course with no pre-requisites.

Next, I brainstormed appropriate teaching approaches/principles to help the students learn the techniques. There are great examples available from statistics education research and suggestions from many educational reports to pull from. For example, I decided to use real data throughout the course, as suggested by the Guidelines for Assessment and Instruction in Statistics Education (GAISE) college report and others (Franklin and Garfield 2006; Singer and Willett 1990). Using real data for this course was an easy decision because it fits very well with the course goals. Real data is messy, and working with it introduces students to the challenges of analyzing real data (Singer and Willett 1990).

I decided to structure the course with a mixture of lectures and labs, including class discussions and group work opportunities, to provide students with a variety of learning methods to help them tackle each topic we were covering. As pointed out in Snee (1993) and suggested in the Curriculum Renewal Across the First Two Years (CRAFTY) statistics report (Moore and Peck and Rossman 2000), students learn in different ways and we need to incorporate a variety of methods into how we teach statistics. Additionally, I knew I would need to use several assessment methods that let me give detailed feedback to students on their work. It is known that students' learning is enhanced when students receive good feedback on the expressions of their ideas (Garfield and Ben-Zvi 2007). For this particular course though, I felt good feedback was essential as students wrapped their minds around the idea that in data analysis, there is often no single correct answer, and that justification of your decisions during analysis is very important.

I also decided to incorporate active learning activities in the course. There has been general enthusiasm about active learning, where students take an active role in their learning in the classroom. There are conflicting accounts of whether or not it is effective for teaching statistics, but numerous studies have found it to be effective (Carlson and Winquist 2011). For a more substantial review and discussion on active learning approaches, see Carlson and Winquist 2011. There are many available active-learning-based resources for teaching introductory statistics, and some efforts have been made to develop resources for more advanced courses (Samsa et al. 2012). Active learning can also be incorporated via cooperative groups (Garfield 1993).

Another decision motivated by the course goals was to spend a solid part of the course on exploratory data analysis and data visualization. Data visualization meant using a lot of graphs and graphics. Its inclusion was motivated in part because the GAISE college report states learning appropriate graph use is an outcome in introductory statistics courses (Franklin and Garfield 2006). Thus, learning appropriate graph use would also be relevant for a course using graphs where students had not previously had introductory statistics. I wanted to keep the focus on students learning what graphs could show, and not become mired in the details about how to create the graphs. This is in line with research about how to get students to view graphs as reasoning tools rather than just "pretty" pictures (Garfield and Ben-Zvi 2007). Graphics can also be helpful teaching tools. For example, Valero-Mora and Ledesma (2011) discuss their experiences teaching some multivariate data analysis techniques using interactive graphics.

In order to have the students generating graphics and develop some computational skills, it was clear that statistical software was needed for the course. The use of statistical software is also recommended by GAISE (Franklin and Garfield 2006). In course design, it is important to consider what level of programming the students can handle. For example, Samsa et al. (2012) discussed student attributes as part of their course development considerations and noted that their students were not skilled programmers. I believed students interested in this course were also not likely to be skilled programmers, but I wanted to introduce them to powerful computing software to give them a basis for developing computational skills. So, I decided to use R with R Commander (Rcmdr) as the primary software for the course (Fox 2005). A variety of other software packages could have been used; see Chance et al. (2007) for other options. When working with any software, educators should note that students may be overwhelmed by software commands. Chance et al. (2007) give guidance for how to avoid this, and I designed the student lab activities with this in mind.

I combined these teaching approaches and ideas, and taught a course using active learning activities, real data, and the R software, to introduce students to exploratory data analysis and multivariate analysis techniques to address the need our students have for these methods. In this case study, I present course design details, an example module (and activities), and outcomes/course evaluation for this applied undergraduate statistics course taught with no pre-requisites. I begin with course design in section two, offering further details on class structure, software, and assessment. Then, as an example module, I present the principal components analysis module from the course in section three. This was the first module for a multivariate analysis technique in the course, and shows how the various course design components were integrated. Finally, in section four, I included some evaluation of the course, including a comparison of student exam performance based on knowledge background, critical comments from students, and how students used their knowledge in other contexts, to see how well I met the course goals. The case study ends with considerations for the future in discussion in section five.

## 2. Course Design

 The final course description summarized what I intended to cover but not how I planned to do it:

"Real world experiments often provide data that consist of many variables. When confronted with a large number of variables, there may be many different directions to proceed, but the direction chosen is ultimately based on the question(s) being asked. In biology, one could ask which observed characteristics distinguish females from males in a given species. In archeology, one could examine how the observed characteristics of pottery relate to their location on the site, look for clusters of similar pottery types, and gain valuable information about the location of markets or religious centers in relation to residential housing. This course will explore how to visualize large data sets and study a variety of methods to analyze them. Methods covered include principal components analysis, factor analysis, classification techniques (discriminant analysis and classification trees) and clustering techniques. This course will feature hands-on data analysis in weekly computer labs, emphasizing application over theory. Four class hours per week."

This section details basic course design considerations – available classrooms, dealing with no pre-requisites, course software, course topics, general module structure, and course assessment.

### 2.1 Classroom Space

Class time each week was divided into four fifty minute blocks which alternated between a lecture classroom and a computer lab classroom. The lecture classroom had a computer projection system with an instructor computer, while the computer lab had a computer for every student. This schedule meant careful planning was necessary to optimize the time in the computer lab where students could try techniques on their own. The computer lab had capacity for 24 students.

### 2.2 Software and Background Knowledge

Students enrolling in the course were not expected to have background statistical or linear algebra knowledge, or any programming experience. The first two weeks of the course were devoted to covering background material to give students a common frame of reference and learning the software.

I planned to use R for the course, due to its power, capacity for graphics, and available packages. However, I was aware that my students were not likely skilled programmers. Instead, I only assumed that the students were digital natives, and would be familiar with computer programs such as Excel. This motivated me to use the graphical interface RCommander (Rcmdr) for R, rather than just the R console (Fox 2005).  The interface for Rcmdr is menu-driven and can be a comfortable starting point for students. For students interested in learning the R command syntax, Rcmdr provides the syntax in a script window. Eventually, all students in the course had to use R script for some of the multivariate techniques. I also used Ggobi to aid in data visualization and exploration (Swayne et al. 2003). Ggobi also interfaces with R via the package Rggobi. The first two weeks in the computer lab were devoted to learning ways of visualizing multivariate data, covering basic univariate and multivariate graphical displays, and using the software.

I covered other background material in the lecture classroom in those two weeks of class. This included some statistics (types of variables, regression basics, ideas on variable selection and transformation), and some linear algebra (matrix notation, scaling effects, eigen decomposition, and singular value decomposition).

**2.3 Course Topics, Principles, and Module Structure**

After the background material in lecture and lab, I planned for the students to tackle the other course topics, listed in Table 1. Students used Lattin, Carroll, and Green's (2003) *Analyzing Multivariate Data* text for readings, which was sufficient for all topics except VI: Classification, which I supplemented with other notes. These topics are all frequently used multivariate data analysis techniques, covered at least partially in most multivariate texts. For some topics, theoretical depth was limited due to having no pre-requisites for the course. Students have natural intuition about some of these techniques (for example, spam filters classify emails as spam or not).

**Table 1**: Course Topics/Modules

| I: Data Exploration and Visualization; Linear Algebra and Statistics Background Material |
|---|
| II: Principal Components Analysis |
| III: Factor Analysis – primarily exploratory, some discussion on confirmatory |
| IV: Multidimensional Scaling – metric and non-metric |
| V: Clustering – hierarchical methods and K-means |
| VI: Classification – linear and quadratic discriminant analysis, nearest neighbor methods, support vector machines, classification trees and random forests |

After the first module, each main course topic received 2-2.5 weeks of class time with between 4-5 lecture days and 3-4 lab days, followed by a homework assignment. The outline of a typical module is provided in Table 2.

I used the lecture days to convey basics about the topic, examples, considerations (such as tuning parameters) and relationships to other topics. The final lecture day was devoted to examining applications of the technique in published work. By showing published articles and having students walk through the analysis with me, students could show what they had learned about the topic so far, and learn to critique statistical work.

**Table 2**: Typical Module Outline

| Prep | Chapter/Sections Assigned from Text |
|---|---|
| Lecture 1 | Motivational Example and Procedure Basics |
| Lab 1 | Explore New Data Set or Motivational Activity |
| Lecture 2 | Procedure Basics and Second Example |
| Lab 2 | Apply New Technique Using R, Class Discussion |
| Lecture 3 | Choosing Tuning Parameters, Connections to Other Topics (e.g. cross-validation procedures, bootstrap) |
| Lab 3 | Further Applications of Technique Using R, Class Discussion |
| Lecture 4 | Application Day |
| Lab 4 | More Applications, Discussion, Homework Assigned |

In between lectures, students explored the techniques in the computer lab. Labs were self/group-paced with instructor availability by request to help with any issues and were written in an active learning style with questions to guide students. I carefully designed the labs and made sure they contained step-by-step instructions for how to conduct the procedures in R/RCommander but had flexibility for students to choose various options relevant to the techniques. The labs also provided a time to explore issues not covered in lecture and for students to learn about the challenges of real data analysis (Singer and Willett 1990). We spent some time discussing outliers and missing values, as well as how the computer software treated those values for each technique.

For each module, I wanted to keep an emphasis on data visualization, use real data for all examples, incorporate active learning activities, and use a variety of learning methods. These key desires were combined in the following ways. Labs always began with data visualization and exploratory analysis, so the students were exploring the data before applying one of the new techniques they were learning. The data sets used for class and lab examples were a mixture of classic and messy data sets, from a variety of disciplines including (but not limited to) biology, music, restaurant management, economics, medicine, chemistry, real estate, sociology, and anthropology. Only one artificial data set was used throughout the entire course, so students were working with real data for every technique[1]. Active learning was incorporated in some lectures where students helped guide the discussion, as well as in most labs, where students worked through examples while addressing various questions. Finally, students had access to a variety of learning methods - readings from a text (Lattin et al. 2003), homework and a course project, class discussions, labs with exploratory components, with opportunities to share with classmates and get summaries from me. I hoped this variety would help students give value to the methods they were learning.

**2.4 Assessment**

Designing assessment for this course was a challenge. Statistical data analysis is not a calculus problem where there is only one correct answer, and I wanted that to be clear to the students. I also wanted to give the students good feedback on the expressions of their ideas, and offer several ways to convey their ideas to me. The breakdown of assessment tools that I decided on for the course is shown in Table 3.

**Table 3:** Assessment Tools

| Tool | # of Assignments/Duration | % of Final Grade |
| --- | --- | --- |
| Homework | 6 – about 1 every 2 weeks, one per module | 20% |
| Midterms | 2 – Concepts I-III, IV-V | 20% each |
| Final Project | Final four weeks of the semester | 10% |
| Take home final | Concepts I-III, V-VI, one week, distributed last day of class | 30% |

Each course concept had a related homework assignment, which was a 4-5 page write-up about analysis of a data set (or two) with supporting work and addressing a few questions I asked. The students got to grapple with homework where there was not necessarily one correct answer, and they had to produce and provide appropriate supporting work. For each assignment, I determined a basic rubric to grade on the statistical content, and while I commented on their writing and made corrections, I did not include a writing assessment on homework. I designed the rubrics so all assignments were a similar point value, giving points for preliminary analysis, appropriate technique choice and discussion of option choice (for example, rotation choice or number of PCs/factors/clusters), as well as addressing any specific questions I had asked.

---

[1] A quick note about that solitary artificial data set: I designed it so that using basic graphing functions (scatterplots) would result in letters being spelled out in the display, which could be properly arranged to spell a word. This was irrelevant for the problem related to the data, but was designed to reinforce the power of data visualization. For details, or that data set and constructed problem, contact the author.

Midterms were held in class during lecture times and consisted of several problems each with multiple parts (with perhaps shared data sets). I supplied all results (graphs and output), so the students were interpreting the results or picking between different options with explanations for their choices, and not applying techniques themselves. I did not write any exam questions where students were given a problem and then would have to suggest an appropriate analysis. Instead, midterms focused on students demonstrating understanding of each technique, reading and interpreting the output. I relied on the take home final and the course project to give me better insight into student abilities to apply appropriate techniques.

Course projects were individual projects where students applied at least two of the main techniques from class on a data set (either supplied by me or a data set they had access to) after brainstorming some questions of interest that they would try to answer. Students had to share their questions with me before delving into their data analysis. They summarized their results in a 10-12 page paper, due shortly before the end of classes.

For the take home final, I combined two open-ended analysis questions (PCA/factor analysis and clustering related) with a classification question where I provided the analysis (due to the significant coding requirement) and asked questions. With this combination of questions, students would have to interpret the results I obtained as well as perform their own analyses. They had a full week to complete this take-home, and I arranged for shorter write-ups (2-3 pages max per question) by not requiring a full exposition of their analysis, just their final choices and rationale for those decisions.

To see how all these course design aspects came together in the course, I present the PCA module in the next section.

## 3. An Example Module - The Principal Components Analysis Module

The PCA module was the first module on a specific multivariate data analysis technique. There were two days of lecture at the start of this module, in order to cover enough material for students to work on their own (or in groups) in lab. Briefly, I present the days in the module stating what happened in class and give some additional comments. I also describe the PCA related assessments used in the course. Many materials described here are provided in the Appendix.

**3.1 PCA Module**

**3.1.0 Prep**

What: Students were assigned to read the PCA chapter from the text.

Comments: One nice feature of the text was a worked example analysis that was presented at the end of each chapter as its own section. This meant students got to see examples in the textbook step by step during the topic development but also as a complete analysis at the end of each chapter.

### 3.1.1 Lecture 1

What: Introduction and motivation of the technique using data students were familiar with from the data visualization module, followed by a simple example using only two variables, and the linear algebra derivation of the technique.

Comments: Students were only required to know that the solution was found by performing an eigen decomposition and did not have to produce a solution by hand.

### 3.1.2 Lecture 2

What: Discussed finer details of method including impact of standardization on solution, number of PCs to retain and why, biplots, and how to interpret the PCs. Two real data examples were used for this discussion. The examples used were a track and field record data set (to show impact of standardization) and a data set on crop production and animal possession in farms in Mali.

Comments: The output for the examples was from R and Rcmdr and was presented in PowerPoint slides for class discussion and comment (see Figure 1 for example discussion slides). This allowed students to see R/Rcmdr output before going in the lab to try the technique themselves.

**Figure 1.** Example Discussion Slides From Lecture 2 from PCA Module

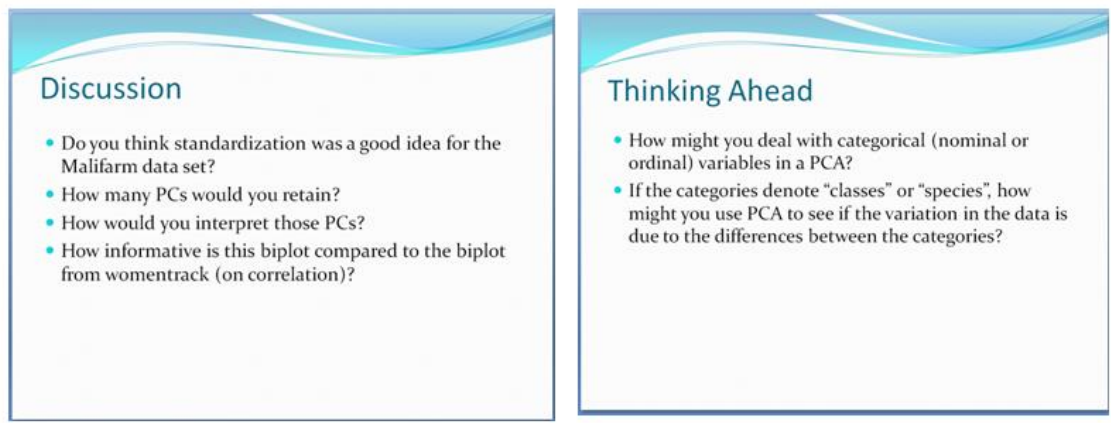

### 3.1.3 Lab 1

What: Introductory lab on PCA using an air pollution data set to examine relationships between pollutants and effect of standardization on the PCA solution. Students performed PCA on the data set using both the correlation and covariance matrix as the input.

Comments: I designed this lab to familiarize the students with performing PCA and reading the output in R/Rcmdr, and it ended with class discussion on the number of PCs to retain. (See Appendix A for PCA lab activities.)

### 3.1.4 Lecture 3

What: (Ungraded) Class quiz, comparison of rules of thumb for retaining PCs, and discussion about choosing the right answer and justifications.

Comments: I had planned to do another example, and discuss ties to other techniques (regression), but that plan was derailed after the quiz. The class wanted to compare the different rules of thumb for how many PCs to keep on other example data sets. This was prompted by one student who asked how the answer was determined for the number of PCs to keep on the quiz, and how they should justify their answers. I explained that in an analysis the statistician has to make the call based on the situation, their judgment and rules of thumb. For the class, I said they should justify it by discussing which rule (or two) they used, and that if I asked them how many PCs a specific rule suggested they retain, we would all get the same answer. For some students, this was likely a first encounter with "more than one right answer" thinking. So we spent considerable time talking about really needing to understand the problem at hand in order to make appropriate decisions about cutoffs, tuning parameters, etc.

### 3.1.5 Lab 2

What: Categorical variables in PCA lab with PCA on a data set on bulls sold at auction with a numerically coded categorical variable. The students explored PCA solutions that included the categorical variable, other solutions that did not include it, and ways to decide if the PC scores provided a separation of groups.

Comments: I designed the lab so students would think about how categorical variables fit or did not fit in a PCA solution. After running the initial PCA with the categorical variable included, they were asked a series of questions about how that categorical variable fit into their solution and if their results made sense. Next, they redid the analysis leaving out the categorical variable to look at the changes. Finally, they started exploring (by plotting the PC scores) whether or not the PCs provided a separation of the groups represented by the categorical variable. Towards the end of class, discussion arose about ordinal variables, indicator variables, discretization of variables, and why it is important to know your data before blindly running a technique. An extra lab was available that had indicator variables in the data set for students who finished the second lab early. (See Appendix A for PCA lab activities.) Most students did not finish Lab 2 during this time.

### 3.1.6 Lecture 4 – Application Day

What: Application day. Four articles were prepared for discussion. The choices for discussion were applications of PCA to reduce the data dimension and provide at least a partial separation of species (bats, frogs, and plants), or an application where PCA was used to reduce the data dimension with interpretability and to classify observations as fraud or not. The students chose the discussion order as bats, fraud, frogs, and then plants. (See Appendix B for article references and details.)

Comments: The final lecture day in each module was our Application Day, where we spent time examining how the technique was used in several publications. During the semester, the application lecture day took a lot of time to prepare. I chose to search/browse through JSTOR and other journal collections, find articles from different disciplines, read/skim the articles myself, and then briefly summarize the research and pull out interesting graphs, methods, and results for the class to discuss. It was a challenge to determine if I had summarized enough of the background material in each article for the resulting statistical application to be of interest to the students. I always prepared several articles (minimum four, usually five, maximum seven) for each course topic, with about equal depth on each, and let the students choose the order we considered them for discussion.

### 3.1.7 Lab 3

What: Continuation of Categorical variables in PCA Lab. Additional lab available involving PCA on hotels and creating new variables (a sum of indicator variables to describe total amenities instead of the indicator variables themselves). Most students were still working with the bulls lab for this lab day.

Comments: At the end of class, we briefly discussed the bulls lab, and I summarized what the extra lab on hotels would have shown them as well (if they wanted to look at it).

### 3.2 PCA Assessment

As a main course topic, PCA had several related assessment materials - a homework and exam questions (midterm and final), and could also have been used for the course projects.

The PCA homework had two associated data sets. For each data set, students had to do some preliminary data exploration, examine the correlation or covariance matrix, perform a PCA, decide the number of components to retain, and explain what their analysis revealed. For example, a student might have argued that three PCs were appropriate to explain their chosen percentage of variation, but that four or five was too many, and supplied a table of proportion of variance explained as support. The PCA homework is shown in Appendix C.

PCA also was a major component on the first midterm and final exam. The first midterm question on PCA can be found in Appendix D.  The question on the take home final exam related to PCA can be found in Appendix E.

Some students used PCA in their course projects. One example data set and set of questions for a course project where PCA was an appropriate technique was a data set on wine cultivars from UCI's Machine Learning Repository (Frank and Asuncion 2010).  It is natural to be curious about how the wines grown by the three cultivars are different, which variables vary the most, and if it is possible to predict which cultivar grew wines if presented with future observations. Several students chose to work with this data set and variations on these questions, and found PCA and various classification techniques were helpful addressing them (the first 2 PCs provide a very nice separation of the wines based on cultivars).

Examples of other data sets students worked with for projects include a data set from a summer baseball internship, a letter recognition (alphabet only) data set, and a data set where the goal was to identify dermatological diseases based on patient attributes. With the course projects, students got to see firsthand how different techniques worked for some data sets and others did not (notably for classification techniques).

## 4. Outcomes and Course Evaluation

Teaching the course with no pre-requisites was a challenge. Did I succeed in making multivariate analysis material accessible to all my students? Is a course with no pre-requisites really viable as an introduction to multivariate data analysis? I had some reflections looking back at aspects of the course. I also wanted to know what the students thought of the course and its aspects. What did they suggest I do differently for future courses? What did they do with their knowledge of the material after the course? This section addresses those questions by looking at student performance in class, my reflections on course aspects, student course evaluations, and short survey replies.

## 4. 1 Student Performance Based on Background Information

To give the reader a sense of the class, Table 4 contains a breakdown of the class student composition by class year, major, and background course (linear algebra or introductory statistics) information. Table 4 also demonstrates that requiring both courses as a pre-requisite would have dropped course enrollment from 19 to 5, and choosing one would have cut it in half.

**Table 4**: Class Student Composition

| Class Year | Major | Background |
|---|---|---|
| 11 seniors | 5 Dual Math/Econ | Group 1 - Linear Algebra and Intro Stat – 5 students |
| 4 juniors | 5 Math | Group 2 – Only Linear Algebra – 5 students |
| 2 sophomores | 4 Econ | Group 3 – Only Intro Stat – 4 students |
| 2 freshmen | 5 Other (Music, Spanish, Psychology, Undeclared) | Group 4 - Neither Linear Algebra or Intro Stat (1 student may have had AP Stat) – 5 students |

It is natural to wonder how students performed in the course given their various backgrounds. Although extensive assessment measures were not undertaken, some analysis of course midterm and final exam scores reveal minor differences in performance of the background groups. Bear in mind the small group sizes and observational nature of this evaluation. All scores reported here were converted into percents. Originally, midterm 1 was out of 60 points, midterm 2 was out of 40 points, and the final exam was out of 75 points.

The midterms were hour long examinations held in class. Midterm 1 covered exploration and data visualization, principal components analysis, and factor analysis in three exam questions with multiple parts. There were no substantial performance differences between groups on midterm 1 (all medians within .5 of each other, with similar IQRs). Midterm 2 scores have an unusual pattern. Midterm 2 covered multidimensional scaling and clustering in two problems with multiple parts each. On this exam, students who had only linear algebra and not introductory statistics, performed poorly relative to their classmates (median 10 percentage points lower), as shown in Figure 2. One student in that group complained that the exam was too long and she left one important (high point value, paragraph answer) question blank, resulting in a low score of 70%.

**Figure 2.** Boxplots of Midterm 2 Scores for the Four Student Groups

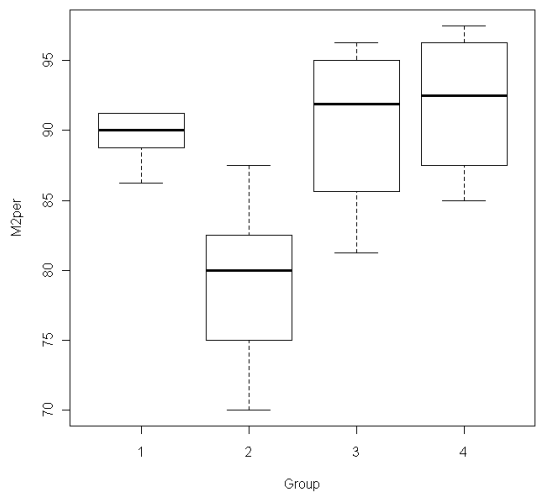

Given the topics covered on midterm 2, I do not see a compelling reason why a linear algebra background would be detrimental to learning the techniques presented. I feel that the poor performance of group 2 on the exam was just an anomaly.

Final exam scores show a predictable pattern, shown in Figure 3. We see that students with a background course did better than those without, and that students with both background courses did a little better than those with just one. The group with no background courses had lower scores, with one outlier who had an exceptional exam.

**Figure 3.** Final Exam Percentages for the Four Student Groups

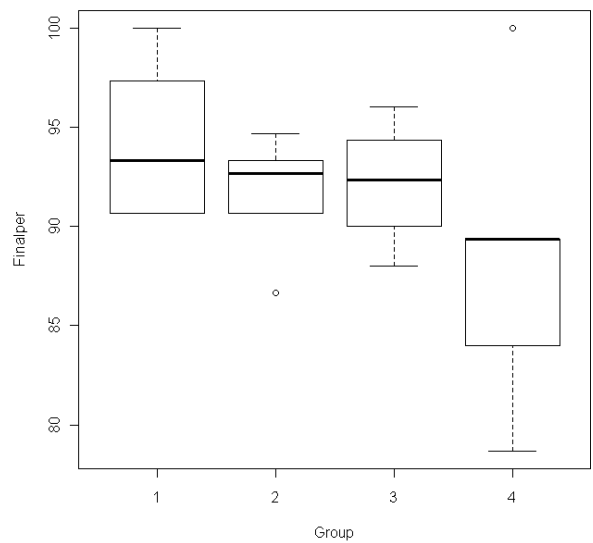

Based on these comparisons, students with a background course (or both) performed comparably or better than their counterparts on the examinations, except midterm 2 for group 2. However, there is not a large gap in performance between students with and without background courses. This suggests the material the students were held accountable for was made accessible to them. I hope to teach the course again, at which time, other appropriate information to compare background groups could be collected.

## 4.2 Some Reflections on the Course

Before looking at student comments, I offer a few reflections on certain course aspects from my perspective – application days, software, and group work in labs.

First, I thought the application days went very well. Class participation in these discussions was high (and not forced through any participation grades). The students had no problems asking why certain procedures had been performed if they were unclear about the choice of procedure.

They also made suggestions about what could have been done differently or improved upon in the analysis. I thought it would be beneficial to students to see how the methods they were studying and results of analyses were explained and presented in several different contexts. I hope this lead to reinforcement of the technique and its uses. Investigating the benefits of activities like this is an area for future research.

For the course software, students were generally receptive to R with RCommander as an interface, especially with instructions on how to obtain all graphs/output, with my help by request at any time. The RCommander (Rcmdr) interface in R was sufficient for all techniques except MDS and classification (Fox 2005). The students grappled with the raw R MDS code, and managed fine, but when it came to classification, and the variety of techniques we studied, I provided the code and analysis. Several students ended up trying classification as part of their course projects, and a few of them came to see me for help with the code, but the rest figured it out on their own, working from two complete coded examples I had given them. Overall, I was very pleased with the students' use of R/Rcmdr and the course techniques. Data was easy to read in and most time was spent exploring the data or analyzing and interpreting results rather than spending hours on data management, though this was included and students gained experience with it more during their projects. I think it was beneficial to keep a lab structure where students were introduced to the software via exploratory visualization analyses early on and used that to start every lab thereafter.

When working on the labs, some students chose to work alone, doing their own work but comparing/discussing results with students or groups around them after they each tried the methods, but most students worked in pairs or a triple (one case) and tried to figure things out as a group. I believe I was very fortunate with this group of students, as they seemed particularly amenable to this structure. In the future though, it might be better to set up cooperative learning groups to help with active learning. I also did not hold them accountable for the group work, or make each group report during the discussions, both components of group work that might be worth incorporating in the future (Garfield 1993).

The student perspective on the various course aspects is explored in the next section.

## 4. 3 Comments on Course Aspects and Related Critical Comments

The students filled out course evaluations before the final exam. The evaluation included questions designed to generate feedback on the course and various course aspects, such as the application days, labs, and software. To elicit further comments and learn what students had done with course material after the course, I recently sent a short survey with questions about the course to the students. I was able to send this additional survey to 18 of the 19 students, and received six replies. In the additional survey, I asked the students if they had used or encountered the techniques elsewhere, how the course had prepared them for that (if they had), to reflect on having used R/Rcmdr, and to offer any critical comments about the course. In this section, I will briefly discuss comments from the evaluations and the short survey, and leave comments about what they did with the course for the next section.

Perhaps the most important course aspect to address is whether or not the material was accessible to all the students, and if they felt that they had gained good understanding of the techniques we had studied, no matter their background. This is related to an evaluation question about the level of theory in the course. Many students were quite content with the applied nature of the course, and felt they got a lot out of the course. They voiced this in their final course evaluations:

> "It's nice to have another math class that is more hands on and has more application rather than theory based courses."
> "I loved it [the course], and I really hope it continues to be open to students without a huge theoretical math background, like me."
> "The material was extremely interesting to me. It was good to take an applied math class that I felt like had real life applications. I am very glad I took this class; it opened my eyes to ways I could use math in my future without just studying theory."

Other students wanted more emphasis on the theory, and there were four critical comments about the level of theory on evaluations. One of those four students pointed out that I presented theory that they were not required to know in depth, and that had been frustrating. This student suggested that I try to find a "happy medium" for teaching the course with more emphasis on the theory, or present less for each technique and cover more techniques. The other three students who made critical comments stated they would have liked to have seen more theory. From the short survey, one student stated that he felt a deeper linear algebra background would have benefited him while learning about the techniques. A second student also thought more linear algebra at the outset of the course would have been beneficial to students who had not had that class. These comments suggest that there is definitely a need to balance the desire of some students for theory while others taking the course might be uncomfortable with theory for courses like this with no pre-requisites. Notably, more of a focus on a linear algebra framework might be necessary at the start of the course, even if students are not required to do matrix operations by hand. Alternatively, linear algebra could be made a pre-requisite for the course, effectively changing the course, and allowing for more theory, but limiting the student audience (which is contrary to my goal of a broad audience).

Another major feature of the course was the application day in each module, and I asked the students what they thought of these days on their course evaluations. Twelve students reported that the application days were beneficial, while three students stated they were not helpful. Here are some comments from students who reported the day as beneficial:

> "I really like the applications where she showed us this stuff in real papers."
> "I also really liked seeing the techniques applied to real studies."
> "The different articles showed us how the material was applied in a different number of [sic] fields and thus appealed to people who were interested in anything from biology to business to restaurant management."

For the students who were critical of these days, two students reported that application days were not helpful to them (one would have preferred an in-depth exploration of one example and one would have rather gone on to other topics for more breadth) and the other reported that because four of the six application lecture days happened to be during our early morning class time (9 am), he did not get much out of them (because he often slept through class).

Given the majority of students held a view that the application day was beneficial, I would keep it in future courses, and as already suggested, try to evaluate how much it helps students learn the various techniques. From the course evaluations, we can see one student remembered the restaurant management example on factor analysis enough to mention it specifically. Did that memory connection help the student learn factor analysis better than a student would have who had not seen a series of examples on application day? This is definitely a topic for future research.

The lab activities were another crucial course aspect where students practiced techniques and learned to code some in R. Over the course of the semester, the lab activities got the students thinking about how data is stored and shared, the importance of appropriate visuals in data analysis, and how vital it is to understand the problem at hand, as well as the techniques being used. The labs had an active-learning style and the students used R with Rcmdr for each one. I was curious to see what students thought about the labs on their evaluations. Some student comments about labs follow:

> "The labs were designed to practice the things we were learning about in class. The labs were also really interesting and always had a point."
> "That I was able to gain some skill in coding and statistical analysis software was actually one of the best parts of the class for me."
> "Lab was essential and very useful in understanding each technique."

There were no critical comments about the content of the labs, but there was one critical comment about R/Rcmdr from the short survey. A student suggested that we should have used Excel or Stata instead of R with Rcmdr because Excel is far more commonly used, and Stata is used by our Economics department. While I do not plan to swap to other software, it might be useful to point out to students that these techniques can be performed in many packages and to spend one day with output from multiple packages so they can see how similar yet different the outputs will be in the context of a single example.

The only other major area where students had many critical comments (7 total) was the choice of textbook. Students commented that they either did not read the textbook or that it was most useful as a supplement after I had presented the material in class. One student pointed out that since it did not use R, it was not very helpful. The number of textbook comments mean that I will consider other texts if the course is taught again. However, based on my previous search for a text for this course, I believe there are few options for good texts where either the theory is presented in a limited way or in "optional" sections for a course with no pre-requisites. Again, a solution to this issue would be to require linear algebra as a pre-requisite for the course, or eventually, preparing my own text/notes for the course.

There were minor comments about timing issues in the course. Three students indicated problems with the course project being too close in time to the final exam and one suggested that perhaps two smaller course projects spaced throughout the course would have been more beneficial than just one project. Besides issues with the project, two students had issues with pacing. One student thought the course was too slow at the beginning and could have covered more topics. Another student would have liked to have spent more time on classification than we did. If the course is taught again, it would be possible to achieve more breadth of techniques within the clustering and classification modules by reducing time on another module.

## 4.4 What Did Students Do with the Knowledge They Gained in the Course?

One of my main motivations for sending the short survey out to course alumnae was to find out how they had used the knowledge they gained in the course. I had limited feedback on this from course evaluations at the end of the semester. In this section, I discuss relevant feedback from their course evaluations and short survey replies.

First, even at the end of the semester when the course was taught, several students had positive things to say about the course and how it had impacted their lives. From their evaluations:

> "Seeing this course on my transcript has been a conversation starter during interviews."
> "It is the one class I have taken at Amherst where I felt that what we were learning was applicable in a large variety of cases, and I could see myself using this material and what I have learned throughout the class at any given point in the future."

One student listed me as a summer internship reference and I was able to state on a phone call that she had in fact used R/Rcmdr to perform a PCA. Some of the seniors were writing theses concurrently with the course that used these techniques, and taking the course helped them to understand what they had been reading about and/or doing themselves, as well as seeing the techniques in a range of applications. Based on personal knowledge of the students and their replies to the survey, at least five of them have gone on to use the techniques in additional coursework or thesis work. Two students reported greatly appreciating learning about data visualization and learning to work with large, real data sets for their current work in the short survey. Another student reported seeing PCA applied often in her current field. Knowledge of R and Rcmdr has proven beneficial for several students (two have entered statistics graduate programs, two reported this knowledge was helpful for job interviews, one has used R since in work, and one reported learning to code was useful for her work with WinSQL), though one reports mostly using Excel for work.

One student reported that the course was essential to her development of a "justify and proceed" attitude for data analysis rather than an attitude that she must search for the ever elusive one "right" answer. She reported that the course homework and lab helped her with this. In a short survey reply, she states:

> "I feel like many quantitatively-minded students at my stage are still looking for a RIGHT answer...and taking this course made me more comfortable with the idea that as long as you can justify your decision as you proceed in the analysis, it's okay, even though there are other decisions that could have been made. Studying with those in my

> program who were so stressed out about finding the RIGHT answer/decision made me really appreciative of that." (Short survey/personal communication)

In introductory statistics courses, this justify and proceed attitude would be difficult to instill in students. However, it is not beyond the purview of other statistics courses.

Finally, the course was important to several students due to its applied nature and course content. Some short survey comments follow:

> "I wish Amherst had more 'practical application of math' courses like yours."
> "I thought that it was one of the best courses I took throughout my Amherst career. It singlehandedly re-sparked my interest in math --which is why I declared the math major upon returning to campus from abroad."

Based on these reports, students used the knowledge gained in the course in a variety of settings (other classes, projects, theses, internships, graduate school, and jobs), and appreciated their undergraduate applied introduction to the topics.

## 5. Discussion

In summary, it is possible to introduce students to multivariate analysis topics outside of an introductory statistics course, without relying on extensive background knowledge of statistics or linear algebra. In this course, a variety of teaching approaches/activities were used to help students tackle the material including active learning, use of real data, and lab activities for exploratory data analysis. Additionally, the students learned some computational skill in R/Rcmdr. The students went on to apply the techniques in other courses, in theses and graduate work, and in their work after Amherst. The students seemed to appreciate the applied nature of the course, and I hope to continue offering it as an applied elective. I plan to teach it at least once more without pre-requisites and to experiment with what theory to present and what to hold students accountable for in that setting. This leads to several areas to consider for the future.

First, there are a few considerations for course content. Students raised a few issues with timing, and several suggested more breadth of techniques. It would be easy to frame multi-dimensional scaling as just one technique in a dimension reduction module, for example. Another idea would be to incorporate more clustering and classification techniques, but what techniques would be best to cut in order to expand those modules? My current thinking is to stay abreast of the multivariate non-regression techniques being applied in thesis projects at the college, and use that knowledge to help guide my decisions regarding content. (Regression is covered by another statistics elective course.)

Next, there are some considerations related to the teaching approaches and techniques used in the course. The group of students in this course was amenable to group work, but I did not use this to its full advantages. I hope to incorporate more of the suggestions from Garfield's work on active learning with cooperative groups in future courses (Garfield 1993). Students might also benefit from more interaction with graphics, such as those explored in Valero-Mora and Ledesma (2011). While students had exposure to interactive graphics with Ggobi, this was also not used to its maximum potential.

Finally, a few areas for future pedagogical research are suggested by this case study. The use of case studies (in a particularly short form during application day) as examples to help students learn techniques bears further investigation.  Additionally, investigating active learning methods in statistics education in courses beyond introductory statistics will likely be enlightening, as we strive to help modern students tackle the statistical concepts that they need to know in the real world.

# Appendix A – PCA Lab Activities

## Lab 2: PCA 1

1 Principal Components in Rcmdr

PCA is one of the multivariate methods included in Rcmdr. It is found under the Dimensional Analysis submenu in the Statistics menu. By default, Rcmdr will output the PC loadings and component variances. You can also ask for a scree plot and can save the PCs to the data set if you plan to use them for later analysis. Additionally, Rcmdr can provide a biplot and the percentage of variation explained by each component with a little bit of help. In this lab, you will investigate a data set on Air Pollution and perform a PCA with the goal of trying to reduce the dimension of the data set. Guides for how to perform the PCA are below, after an outline for your preliminary data analysis.

2 Air Pollution Data Set (Johnson/Wichern)
The data on air pollution is a series of measurements at noon in the LA area. The variables present in the data set are wind, solar radiation (solar), CO, NO, NO2, O3, and HC (hydrocarbons). To learn a bit more about some of these and their role in air pollution, steer your web browsers to http://www.epa.gov/air/urbanair/. This should open a page from the EPA on six common air pollutants. Take a few minutes to skim through what the EPA says about each pollutant that we have in our data set. Then, answer the following questions.
1. What variables are present in our data set but not listed on the EPA page? Are these variables also pollutants or are they included for some other reason (i.e. extraneous, possibly explanatory)?

2. Based on your readings, are the pollutants listed on the EPA page independent or would you expect them to be correlated?

3. Are there any other variables (pollutants or extraneous) that your reading suggests you might have wanted included in this data set?

2.1 Preliminary Analysis
Perform a preliminary analysis of the air pollution data (available from the text file online). You should be jotting down notes/comments about each variable (what type of variable it is, shapes of distributions, outliers, descriptive statistics, as appropriate), then doing bivariate analysis scatterplots and correlations) as appropriate.

Wind:

Solar radiation:

CO:

NO:

NO2:

O3:

HC:

Bivariate findings:

At this point, you should want to stop and consider what you have found. Among the questions to think about you might ask: were there any points you think are outliers that should be removed? Why? (This is where you can catch data entry errors). In the event you do consider you have outliers, you can run the PCA with them once and then remove them and run it again to see the impact they have on your analysis. Observations can be left out temporarily assuming you have the case number using the subset cases command inside PCA. For example, to remove observations 1, 3, and 7, temporarily from the analysis you would enter -c(1,3,7) in the subset expression box.

This lab is designed to teach you how to perform PCA in R/Rcmdr, but you should bear in mind what PCA can be used for. Recall from class that one use is to create new variables that are uncorrelated when your original data set has several highly correlated variables as a means of dimension reduction. Does it appear based on your preliminary analysis that a PCA on this data might be useful for dimension reduction in this fashion? Explain.

# 3 PCA
## 3.1 Performing a PCA

We turn our attention now to performing the actual PCA. The basic PCA output for the data set using all variables and using the covariance matrix in the PCA is obtained as follows:
1. From the Statistics menu, select Dimensional analysis and then select Principal components analysis.
2. In the window that opens, select all variables.
3. Since we want the PCA on the covariance, uncheck the option for correlation matrix.
4. Check the option for the scree plot.
5. Click Ok. (Adjust subset expression if desired).

This generates the default Rcmdr PCA output. For the additional output we want you need to:
1. In the script window (in Rcmdr), find the line that begins with .PC and ends with data=     ), where the blank has been filled with the data set name you used.
2. Highlight this line (in some cases, it might be more than one line).
3. Click Submit. You will not see anything happen directly, but it will have rerun the PCA and has the results active.
4. Back in the script window, type the following:
biplot(.PC)
summary(.PC)
loadings(.PC)
5. Highlight each one in turn and hit the submit button (or all three at once). The first one will provide a biplot for the PCA. The second gives the percentage of variation explained by each component and the third gives another variation of the loadings output.
6. If you want the screeplot back, simply highlight screeplot(.PC) in the script window and hit Submit. In this way you can go back and forth between the graphs you generated.
7. (On Correlation Matrix) To obtain a PCA on the correlation matrix, simply do not uncheck the correlation matrix option in step 3 in the default PCA output section.

## 3.2 PCA on Covariance Results

Look through your PCA output on the covariance and address the following questions:
1. How many PCs were extracted in total? Why?

2. How many PCs would you choose to keep and why? Discuss 2 different ways of making this decision. (Final choice is subjective).

3. What variable appears to be dominating the PCA based on the biplot?

4. Compare the default loadings display with the display you obtained from loadings(.PC). What are the differences between these displays?

5. Do you think interpreting the PCs from the covariance matrix is helpful in understanding the variability in the data?

3.3 PCA on Correlation Results
Now, obtain a PCA based on the correlation matrix, and address the following questions:
1. How many PCs would you choose to keep and why? Discuss 2 different ways of making this decision. (Final choice is subjective).

2. Report the PC loadings for the PCs you choose to keep (your choice of loadings output).

3. Provide possible interpretations for each PC you choose to keep.

4. Take a look at the biplot. What relationships do you see between the variables? For example, what variables are higher levels of solar radiation associated with?

5. How informative is this biplot in understanding what the relationships are between the variables? How did you make this determination?

3.4 A PCA of Your Choice

Perform a PCA on a subset of the variables you are interested in (for example, you might pick just the pollutants, or a certain pollutant and wind/solar, etc.). You may choose to perform the PCA on either the correlation or covariance matrix. (Alternatively, you could remove some outliers and rerun the original PCA, or do that and select different variables.) For your PCA, report the variables you selected, whether you chose the covariance or correlation matrix, the number of PCs you would retain, possible interpretations for those PCs, any interesting relationships shown by your biplot, and how this analysis compares to the previous ones. Class will discuss so bring thoughts/comments.

3.5 Think About

What sort of PCs might be generated for a data set where the strongest (magnitude) starting correlation between any two variables was .1?

Suppose you ran a PCA on a data set where all variables were measurements in inches and another student ran a PCA on the same data set but had transformed all the variables to be in centimeters. What differences would you expect between the PCA solutions, biplots, etc.?

**Lab 3 and 3.5: PCA 2 and 3**

In this lab, you will be investigating the impact of categorical variables (and their coding) on PCA with a data set on bulls sold at auctions, as well as investigating a data set on hotels in the D.C. area and creating a new variable to examine the relationship between costs, hotel size, and amenities. This will give you more experience with PCA as well as statistical concepts like handling categorical variables and creating new variables from indicator variables.

1 Bulls Data Set
This data set contains information on bulls sold at auction and is online as bulls.txt. The Ggobi format is also available with breed color/symbol coded (via the copy/paste and resave as .xml file method). The data is courtesy of Mark Ellersieck and is taken from Johnson/Wichern. The variables in the data set include:
breed = 1= Angus; 5= Hereford; 8 = Simental
salepr = sale price at auction
yrhgt = yearling height at shoulder (inches)
ftfrbody = fat free body (pounds)
prct_b = percent fat-free body
frame = scaled from 1-8 (1 = small, 8 = large)
bkfat = back fat (inches)
saleht = sale height at shoulder (inches)
salewt = sale weight (pounds)

Suppose we want to know what the major differences between breeds are based on the bulls sold at auction. Could a PCA possibly help us with this?

What about if we wanted to obtain a single numerical value as a summary score for each bull to use to set a sale price instead of using an auction sale format for future bulls? Could PCA help with this? Explain.

What variables are nominal? What variables are ordinal? What variables are ratio-scaled?

Are the nominal variables going to be recognized as nominal by Rcmdr? Why or why not?

For the purposes of this exercise, you will need to convert breed into a factor variable so that Rcmdr knows it is categorical. However, since we want to examine the impact of a categorical variable, you will need to save it with a new name so we can still access the original variable. To do this, follow these steps:
1. In the Data Menu, select Manage variables in active data set, and then select Convert numeric variables to factors.
2. Select breed.
3. Enter the new name as breed2 (so you won't replace breed).
4. Leave supply level names (since you actually have the breed info above).
5. Click Ok.
6. In the sub window, enter in that 1=Angus, 5=Hereford, and 8=Simental.
7. Click Ok to finish.
8. View the data set to verify that breed2 has been added.

1.1 Preliminary Analysis

To obtain a basic understanding of the data, perform a preliminary analysis. Note that you have a categorical variable this time (breed2) so you may make more interesting bivariate comparisons when you get to that stage (e.g. side-by-side boxplots). Again, you should jot down some comments/notes about each variable and then your bivariate findings, including a look at correlations. You can do this quickly - recommend scatterplot matrices with histograms and boxplots on diagonal for quick looks.

Breed/breed2:

Yrhgt:

Ftfrbody:

Prct_b:

Frame:

Bkfat:

Saleht:

Salewt:

Salepr:

Bivariate findings:

Any possible outliers you want to temporarily remove from analysis to assess their impact?

1.2 PCA with Breed
Perform a PCA on the correlation matrix with all variables selected (removing observations is up to you). You will notice that breed2 is not an option since it is recognized as categorical by Rcmdr, but we still have breed included so the information is there. The end result is a PCA with a categorical variable included by giving it a numerical coding. Obtain the PC loadings, screeplot, and biplot. Save both the screeplot and biplot by copying/pasting into a Word document to compare to later. Then answer the following questions:

1. Breed is coded as 1,5,8 numerically in order to be included in the PCA. Do you think using 1,2,3 would get different results? How would this coding impact results? What about coding as 8,5,1 or 5,1,8?

2. How many PCs would you retain? Why? Discuss at least 2 methods for determining this.

3. Provide interpretations for the PCs you selected to keep.

4. Do you think the biplot is very informative? Why?

5. How does breed fit into the PCs you kept?

6. How does breed relate to the other variables based on the biplot?

Do not worry if your answers to 5/6 are not satisfactory or seem odd to you. They should! For example, we can't really compute a correlation between breed and the other variables but R does with the numerical coding we gave it for breed. Generally, PCA is not done with categorical variables unless they are indicator variables (just 0/1) or ordinal. The additional ordering information from ordinal variables can help, but even still standardizing and correlations do not really make sense. However, statisticians will often look to see if a PCA can capture the differences between levels of a categorical variable, which is what we will try next. The procedure that turns out to be nearly equivalent to a PCA for categorical variables is called correspondence analysis.

1.3 PCA without Breed
Perform a PCA on the correlation matrix with all variables selected except breed. Obtain the PC loadings, screeplot, and biplot. Save both the screeplot and biplot by copying/pasting into a Word document to compare to the ones from above. Then answer the following questions:

1. How many PCs would you retain? Why? Discuss at least 2 methods for determining this.

2. Provide interpretations for the PCs you selected to keep.

3. Do you think the biplot is very informative? Why?

4. Do the PCs appear to be similar even without the breed information? How can you tell?

5. Rerun this analysis and this time, save the PCs (these are the PC scores) to the data set that you have decided to keep (it will prompt you to tell it how many to save - which means you need to have already run it once to figure that out, unless you just save them all the first time through). They will appear as variables PC1, PC2, etc. in your data set. Look at scatterplots (or a scatterplot matrix) for the PCs. Should you see any overall association in the scatterplots?

6. Verify that the correlations between PCs are 0.

7. Plot the scatterplot (or scatterplot matrix) for the PCs by groups using breed2. Is it more clear now whether the directions of maximum variation correspond to the differences between breeds?

8. Discuss whether or not you think this PCA captured some information that could help distinguish between the breeds and how helpful you think this PCA is in trying to understand differences between the breeds.

9. Recall that the loadings F from class which are correlations between the PCs and original variables are equal to the R reported loadings times the square root of the associated eigenvalue (which can be found as the component variance in the output).
a. What PC explains most of the variation in bkfat? Can you tell without the eigenvalue information?

b. What 3 variables would you expect PC2 to be most strongly correlated (positive or negatively) with? Verify by obtaining the correlations with Rcmdr (you may have to add more PCs to the data set) or doing computations.

c. What 4 variables would you expect PC2 to be least strongly correlated with? Verify.

d. What are the reported loadings for the 2 variables most strongly loaded on PC1? What are the corresponding correlations?

e. Look at the loading of prct_b on PC1 and PC2. On which PC does it have a higher loading?

f. Look at (or compute) the correlation of prct_b with PC1 and PC2. With which PC does it have a higher correlation? (This can happen because of the eigenvalues associated with each PC!)

g. The first eigenvalue is roughly 4.25 based on the scree plot. Provide its exact value and compute the loading for yrhgt on PC1 using the R loading and the eigenvalue. Does it match the correlation?

2 Hotels Data Set

This exercise is available if you have completed everything else. Be sure you understand the computations in 9 above.

Hotel managers want to know how their hotels differ in terms of amenities offered as well as prices, so they collected a data set (this is self-reported data). The data is from Choice hotels in the D.C. area (Johnson 1998). The variables are:
location (abbreviated)
number of rooms
min and max summer rates in 1995 for 2 people
indicator variables for dining room (dining), cocktail lounge (lounge), free continental breakfast (bkfst), and pool (yes = 1, no =0)
type of hotel - CI for comfort inn, QI for quality inn and CH for clarion hotel

Identify the type of each variable.

Note that some of the categorical variables are numerically coded. For the purposes of this exercise, you will want to leave the four indicator variables as 0/1 responses. However, you may want to create categorical versions of them (recommend naming them as diningC or similar to distinguish) for the preliminary analysis.

2.1 Preliminary Analysis
To obtain a basic understanding of the data, perform a preliminary analysis. Note that you have several categorical variables this time so you may make more interesting bivariate comparisons when you get to that stage (side-by-side boxplots). Again, you should jot down some comments/notes about each variable and then your bivariate findings.

Number of rooms:

Min rate:

Max rate:

Dining:

Lounge:

Bkfst:

Pool:

Type:

Bivariate findings:

2.2 PCA with Indicators

Remember that the hotel managers want to know how their hotels differ in terms of amenities offered and prices. Perform a PCA on the correlation matrix with all numeric variables selected (this includes the four indicators). Obtain the PC loadings, screeplot, and biplot. Save both the screeplot and biplot by copying/pasting into a Word document to compare to later. Then answer the following questions:

1. How many PCs would you retain? Why? Discuss at least 2 methods for determining this.

2. Provide interpretations for the PCs you selected to keep.

3. Do you think the biplot is very informative? Why?

4. Does it appear that more expensive hotels have more amenities? How about larger hotels? Which specific amenities seem to be related to the more expensive hotels or the larger hotels? How can you tell?

2.3 PCA with Amenities

Remember that you can add, remove, transform, and create new variables if you think it helps with your solution, with meeting assumptions, etc. Rather than use four indicator variables, perhaps just one variable "amenities" will do a better job portraying the level of amenities at the hotels.

Create "amenities" from under the Data Menu, by selecting Manage variables in active data set and then Compute new variable. Compute it as the sum of the four indicator variables. Do a brief analysis of this new variable and then perform a PCA on the correlation matrix with all variables selected except the four indicator variables. Obtain the PC loadings, screeplot, and biplot. Save both the screeplot and biplot by copying/pasting into a Word document to compare to the ones from above. Then answer the following questions:

1. What does the brief univariate analysis of amenities reveal? How about a scatterplot for amenities vs. each of the rates? Amenities vs. number of rooms?

2. How many PCs would you retain? Why? Discuss at least 2 methods for determining this.

3. Provide interpretations for the PCs you selected to keep.

4. What does the biplot reveal to you about the relationships between costs and amenities? Between number of rooms and amenities? How useful is the biplot?

5. How would you help the hotel managers describe the types of differences between hotels?

**Appendix B** - Article Resources for Topic Exploration in the PCA Application Day

**Articles:**

Agyeman, V., Swaine, M., and Thompson, J. (1999), "Responses of Tropical Forest Tree Seedlings to Irradiance and the Derivation of a Light Response Index," Journal of Ecology, 87 (5), 815-827.

Birch, J. (1997), "Comparing Wing Shape of Bats: The Merits of Principal-Components Analysis and Relative- Warp Analysis," Journal of Mammalogy, 78 (4), 1187-1198.

Brockett, P., Derrig, R., Golden, L., Levine, A., and Alpert, M. (2002), " Fraud Classification Using Principal Component Analysis of RIDITs," The Journal of Risk and Insurance, 69 (3), 341-371.

Fernandez de la Reguera, P. (1987), "Identifying Species in the Chilean Frogs by Principal Components Analysis," Herpetologica, 43 (2), 173-177.

**Article Summaries:**
Briefly, the first article I prepared was an old (1987) article about Chilean frogs where a PCA on 11 measurement variables resulted in the first two principal components explaining 86% of the variation and providing reasonable separation of the four frog species into 3 groups (two species were intermixed on the first 2 PCs) (Fernandez de la Reguera 1987). The second article was similar, but more recent, and focused on characteristics of bat wings for four different species of bats. The article contained several applications of PCA on different bat-related data sets, but focused on PC interpretability and whether the species were separated when viewed using PC axes (Birch 1997). The third article I prepared was on a designed experiment comparing tropical tree seedling growth performance in different amounts of shade. The first two PCs in this example only accounted for 56% of the variation, and the second PC separated three drought tolerant species from the rest (Agyeman, Swaine and Thompson 1999). The final article I prepared for the class was a more complex article using PCA to help detect fraud. The article placed emphasis on the first PC score as being a one-dimensional summary measure and interpreting the PC. However, it did not have as many easily accessible graphs/descriptions as the other articles (Brockett et al. 2002). The students chose the discussion order as bats, fraud, frogs, and then plants.

**Appendix C**: PCA Homework Assignment

Two Data sets. First is designed to be shorter and is just a check on effects of standardization. Second is open-ended.

Q1: Investigate scaling phenomena in principal components by performing a principal components analysis on the *mentrack.txt* data set on both the original variables and standardized variables (i.e. the covariance and correlation matrices). How many PCs would you keep from each analysis? How do the analyses compare? Are the effects of scaling here the same as those seen in the *womentrack.txt* analysis? (Remember the PowerPoint with the results from this is online, and I will post the womentrack data set just in case you want to look directly at it also.) *mentrack.txt* contains men's track records for many countries from 2005 from the IAAF/ATFS Track and Field Handbook. For variables, country is recorded and then the following records:
100m, 200m, 400m all recorded in seconds;
800m, 1500m, 5000m, 10000m and the marathon all recorded in minutes;
marathon is 26.2 miles or 42195 meters for possible conversions.

Note: If you want the biplot to use the country names, all you need to do is:
Data>Active Data Set>Set Case Names, and then select country as the variable containing the row names.

Q2: Osteoporosis is a bone disease characterized by a decrease in bone mineral density which results in an increased fracture risk, typically discussed as a disease affecting older women. *bonemineral.txt* contains bone mineral content in dominant radius (dradius) and radius, dominant humerus (dhumerus) and humerus, and dominant ulna (dulna) and ulna in older women. These measurements were taken *before* a study which was investigating the effect of exercise and dietary supplements on slowing bone loss (data courtesy of Everett Smith, from Johnson and Wichern). Researchers want to know which arm and which bones to focus on in their analysis *after* the study when they plan to retake at least some of the measurements. Perform an appropriate analysis on the *bonemineral.txt* data set to help the researchers. You should include a discussion of subjective decisions/cutoffs you make and interpretations of your results.

**Appendix D:** PCA Question on Midterm 1 – Question Spacing Eliminated

Questions 2 and 3 both use the olive data set.

The olive data consists of the percentage composition of 8 fatty acids (palmitic, palmitoleic, stearic, oleic, linoleic, arachidic, linolenic, eicosenoic) found in the lipid fraction of 572 Italian olive oils, along with some other categorical variables. The percentage composition variables are all metric. A junior researcher decides to explore the data set and then run a PCA and factor analysis using all the percentage composition variables in an attempt to uncover interesting data features.

3. The junior researcher also performed a PCA on the olive data set. The researcher remembered something about PCA on correlation usually being a better choice than covariance, so the PCA output below comes from PCA on the correlation matrix.

a. Based on the descriptive statistics, what would have happened if the researcher had run the PCA on the covariance? What would the PCA have shown?

| Fatty Acid | Mean | SD |
|---|---|---|
| Arachidic | 58.10 | 22.03 |
| Eicosenoic | 16.28 | 14.08 |
| Linoleic | 980.53 | 242.80 |
| Linolenic | 31.89 | 12.97 |
| Oleic | 7311.75 | 405.81 |
| Palmitic | 1231.74 | 168.59 |
| Palmitoleic | 126.09 | 52.49 |
| Stearic | 228.87 | 36.74 |

b. The researcher takes a quick look at the correlation matrix. The correlations range from near 0 to over .8 in magnitude. Is PCA appropriate to run on the data?

**Yes    No**    (Circle one, no explanation).

Here's most of the PCA on correlation output:

```
unclass(loadings(.PC))   # component loadings
              Comp.1      Comp.2      Comp.3     Comp.4      Comp.5
Arachidic   -0.2283036  0.44719396 -0.42664494  0.4816544   0.14659527
Eicosenoic  -0.3118678  0.40476916  0.30085585 -0.3322221  -0.67153429
Linoleic    -0.3656954 -0.34339930 -0.08747773  0.5124909  -0.40127538
Linolenic   -0.2189871  0.60483760 -0.19103316 -0.0988132   0.12507081
Oleic        0.4941749  0.15866175 -0.08011486 -0.2001074  -0.01552215
Palmitic    -0.4607435 -0.04958406  0.11445834 -0.2804312   0.53473943
Palmitoleic -0.4502258 -0.24090732  0.14260264 -0.2118225   0.13841908
Stearic      0.0986447  0.25837844  0.80215910  0.4708217   0.21340068
              Comp.6      Comp.7     Comp.8
Arachidic    0.55365142  0.06527504 0.03996552
Eicosenoic   0.25657629 -0.13959613 0.04168750
Linoleic    -0.30497855 -0.07768793 0.46687817
Linolenic   -0.69784174  0.19096065 0.02943890
Oleic        0.11309403  0.18074878 0.79903372
Palmitic     0.07699892 -0.52540418 0.35438653
Palmitoleic  0.16728954  0.78680816 0.08856309
Stearic     -0.03064009  0.07722664 0.07703841

.PC$sd^2    # component variances
     Comp.1      Comp.2      Comp.3      Comp.4      Comp.5      Comp.6
3.721410009 1.765797520 1.016355435 0.792898832 0.333817667 0.248818666
```

```
     Comp.7       Comp.8
0.118820109 0.002081762

summary(.PC) # proportions of variance
Importance of components:
                          Comp.1    Comp.2    Comp.3    Comp.4     Comp.5
Standard deviation     1.9290956 1.3288331 1.0081446 0.89044867 0.57776956
Proportion of Variance ????????? 0.2207247 0.1270444 0.09911235 0.04172721
Cumulative Proportion  ????????? 0.6859009 0.8129454 0.91205772 0.95378493
```

c. What percentage of variation did the first component explain?
d. How many PCs would you keep if you were using Kaiser's rule?
e. What was the correlation between the $1^{st}$ PC and Oleic (variable 5)?
f. What percentage of the variation of Stearic (variable 8) is explained by the third PC?
g. Compare your chosen 3 factor FA solution (from question 2) to the first 3 PCs. What similarities and differences do you see?
h. In general, describe three differences between PCA and FA. How are the methods different?

It turns out that the "additional categorical variables" mentioned that are also part of this data set are Region (3 regions) and Area (9 areas – 4,2,3 from the regions). Region is coded 1-3, and area is coded 1-9. The regions do move north to south, but the areas within are not organized in any particular fashion for coding.

i. Explain why you wouldn't run a PCA with Area entered as a quantitative variable.

j. The junior researcher planned to make side-by-side boxplots (not shown) to see if the first PC could distinguish between the regions and areas. What would you expect to see in the plots if the first PC could be used to separate the categories of either variable?

k. It turns out the first PC alone is insufficient to separate the regions (or areas). Here is plot of the scores symbol coded and colored by region. Comment on whether the first 2 PCs considered together seem to be able to separate the regions (color picture passed out separately).

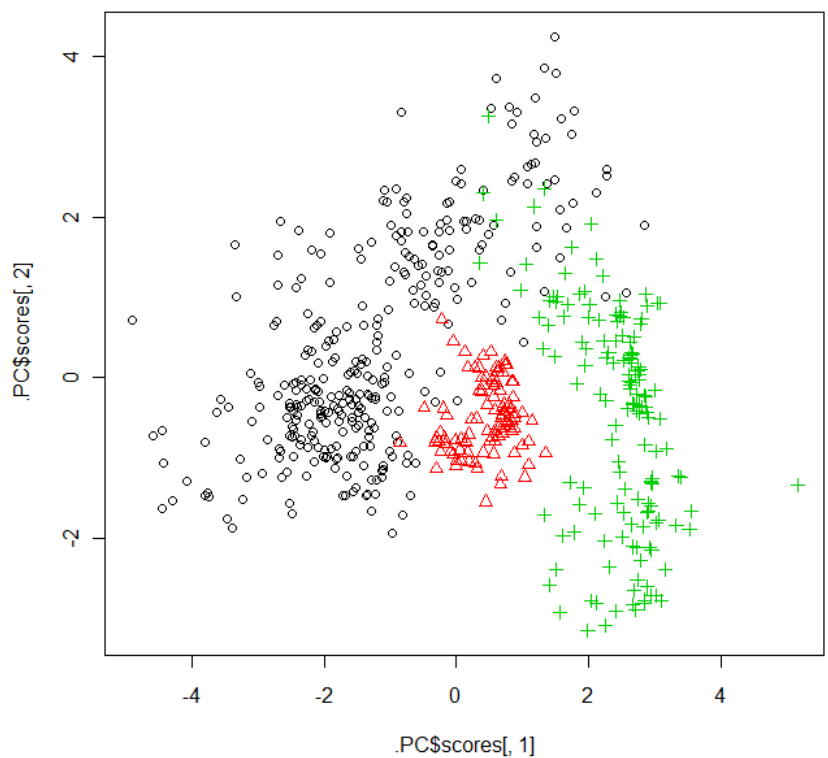

**Appendix E:** PCA Related Take Home Final Question

Question 3. A Gaming Experience

A computer science student has written a portion of a video game while studying the behavior of some computer algorithms. For a particular encounter in the video game, ten different inputs are needed to determine the outcome of the encounter – i.e. whether the player wins or loses a fight against a monster. A set of 190 instances of the encounter have had their inputs recorded, but outcome is not recorded for you. The data set is available as *csgame.txt*. The inputs are: BaseAttack, BaseDefense, Charisma, Constitution, Dexterity, Intelligence, MagicDefense, MonsterAttack, MonsterMagic, and ZoneLevel. The first 7 inputs are player attributes and higher values denote a stronger player character. You should note that BaseDefense is based on armor, and MagicDefense is not, and also that it is usually the case that characters in plate armor (the strongest usually available in these sorts of games) are usually not strong magic users. The $8^{th}$ and $9^{th}$ variables are monster attributes where higher values indicate a stronger, more powerful monster. Finally, ZoneLevel is an indication of how hard the zone where the encounter takes place is (higher values should be a harder zone). The CS student is still working out some issues with ZoneLevel, so it is not well incorporated into the current program. Some of the input names are based on concepts from Dungeons and Dragons.

***Examine the data set and perform a relevant analysis to determine how the variables are related, whether you can somehow reduce the number of input variables using the relationships between variables, and whether or not there seem to be underlying features shared by the input variables. Provide relevant interpretations of your findings if you find results you deem significant.*** Note that these 190 instances were test cases for the encounter so their input values were provided by the CS student, and that you *may* be able to determine how the input values were chosen. You should summarize your findings in a 2-3 page write-up with some supporting work.

**References:**

Brown, E. and Kass, R. (2009), "What is Statistics?," The American Statistician, 63 (2), 105-110.

Carlson, K. and Winquist, J. (2011), "Evaluating an Active Learning Approach to Teaching Introductory Statistics: A Classroom Workbook Approach," Journal of Statistics Education [Online], 19 (1).

Chance, B., Ben-Zvi, D., Garfield, J., and Medina, E. (2007), "The Role of Technology in Improving Student Learning of Statistics," Technology Innovations in Statistics Education [Online], 1 (1), http://escholarship.org/uc/item/8sd2t4rr

Cobb, G. (2007), "The Introductory Statistics Course: A Ptolemaic Curriculum?," Technology Innovations in Statistics Education [Online], 1 (1), http://escholarship.org/uc/item/6hb3k0nz

Fox, J. (2005), "The R Commander: A Basic-Statistics Graphical User Interface to R," Journal of Statistical Software, 14 (9).

Frank, A., and Asuncion, A. (2010), UCI Machine Learning Repository. Irvine, CA: University of California, School of Information and Computer Science. Available online at: http://archive.ics.uci.edu/ml/index.html

Franklin, C., and Garfield, J. (2006), "The Guidelines for Assessment and Instruction in Statistics Education (GAISE) Project: Developing Statistics Education Guidelines for Pre K-12 and College Courses," In G.F. Burrill, (Ed.), Thinking and Reasoning about Data and Chance: Sixty-eighth NCTM Yearbook (pp. 345-375). Reston, VA: National Council of Teachers of Mathematics. Available from: http://www.amstat.org/Education/gaise/GAISECollege.htm

Garfield, J. (1993), "Teaching Statistics Using Small-Group Cooperative Learning," Journal of Statistics Education [Online], 1 (1).

Garfield, J., and Ben-Zvi, D. (2007) "How Students Learn Statistics Revisited: A Current Review of Research on Teaching and Learning Statistics," International Statistical Review, 75 (3), 372-396.

Gould, R. (2010) "Statistics and the Modern Student," International Statistics Review, 78 (2), 297-315.

Kuiper, S. and Sklar, J. (2012), Practicing Statistics: Guided Investigations for the Second Course, Boston: Pearson Education, Inc.

Lattin, J., Carroll, J., and Green, P. (2003), Analyzing Multivariate Data, Cengage Learning, Brooks/Cole.

Moore, T., Peck, R., and Rossman, A. (2000), "Statistics: CRAFTY Curriculum Foundations Project."  Chapter 14 in Curriculum Foundations Project: Voices of the Partner Disciplines from the Committee for the Undergraduate Program in Mathematics (CUPM). Available online from http://www.maa.org/cupm/crafty/Chapt14.pdf

Nolan, D. and Temple Lang, D. (2010), "Computing in the Statistics Curricula," The American Statistician, 64 (2), 97-107.

Samsa, G., Thomas, L., Lee, L., and Neal, E. (2012), "An Active Learning Approach to Teach Advanced Multi-predictor Modeling Concepts to Clinicians," Journal of Statistics Education [Online], 20 (1).

Singer, J., and Willett, J. (1990), "Improving the Teaching of Applied Statistics: Putting the Data Back Into Data Analysis," The American Statistician, 44 (3), 223-230.

Snee, R. (1993), "What's Missing in Statistical Education?," The American Statistician, 47 (2), 149-154.

Swayne, D., Temple Lang, D., Buja, A., and Cook, D. (2003), "Ggobi: Evolving from XGobi into an Extensible Framework for Interactive Data Visualization," Computational Statistics and Data Analysis, 43 (4), 423-444.

Tintle, N., VanderStoep, J., Holmes, V., Quisenberry, B., and Swanson, T. (2011), "Development and Assessment of a Preliminary Randomization-Based Introductory Statistics Curriculum," Journal of Statistics Education [Online], 19 (1).

Valero-Mora, P. and Ledesma, R. (2011), "Using Interactive Graphics to Teach Multivariate Data Analysis to Psychology Students," Journal of Statistics Education [Online], 19 (1).